\documentclass[12pt,a4paper,twocolumn,reprint,superscriptaddress]{revtex4-1}

\usepackage{dcolumn}
\usepackage{bm}

\usepackage{amsmath}
\usepackage{amssymb}
\usepackage{color}
\usepackage[usenames,dvipsnames,svgnames,table]{xcolor}
\usepackage[table]{xcolor}
\usepackage[english]{babel}
\usepackage{graphicx}
\usepackage{array}
\usepackage{multirow}
\usepackage{natbib}

\newcommand{\avg}[1]{\ensuremath{\left<{#1}\right>}}

\graphicspath{ {./figures/} }

\begin{document}

\author{Bingqing Cheng}
\affiliation{Laboratory of Computational Science and Modeling, {\'E}cole Polytechnique F{\'e}d{\'e}rale de Lausanne, 1015 Lausanne, Switzerland}

\author{Michele Ceriotti} 
\affiliation{Laboratory of Computational Science and Modeling, {\'E}cole Polytechnique F{\'e}d{\'e}rale de Lausanne, 1015 Lausanne, Switzerland}
\email{michele.ceriotti@epfl.ch}

\title{Direct path integral estimators for isotope fractionation ratios} 


\begin{abstract}
Fractionation of isotopes among distinct molecules or phases is a quantum effect
which is often exploited to obtain insights on reaction mechanisms, biochemical, 
geochemical and atmospheric phenomena. 
Accurate evaluation of isotope ratios in atomistic simulations is challenging, because one needs to 
perform a thermodynamic integration with respect to the isotope mass, along with 
time-consuming path integral calculations. 
By re-formulating the problem as a particle exchange in the ring polymer 
partition function, we derive new estimators giving direct access to the differential 
partitioning of isotopes, which can simplify the calculations by avoiding thermodynamic integration. 
We demonstrate the efficiency of these estimators by applying them to investigate the isotope 
fractionation ratios in the gas-phase Zundel cation, and in a few simple hydrocarbons. 
\end{abstract}

\maketitle

\section{Introduction}

There is little doubt that modelling explicitly the behavior of 
electrons requires a quantum mechanical treatment. The nuclei of atoms, on the other hand, are typically treated
as classical point particles in atomistic simulations. In many cases -- most notably for the lighter
elements -- the quantum nature of the nuclei gives rise to significant physical effects even
at room temperature or above. Examples are the large deviations from the Dulong-Petit limit of
the heat capacity of solids, the large isotope kinetic effects that are often observed in reactions that 
involve a hydrogen transfer as the rate-limiting step~\cite{liu1993direct}, and the deviations of the particle momentum distribution
of nuclei from the Maxwell-Boltzmann predictions~\cite{andr+05advp}. 

One of the most evident signs of the quantum nature of light nuclei is the fractionation
of two isotopes $\text{X}$ and $\text{X}'$ between two environments $\text{A}$ and 
$\text{B}$ -- where the two environments could correspond to different thermodynamic 
phases, different compounds or inequivalent positions in the same molecule
\begin{equation}
   \text{X}_{\text{A}}+\text{X}'_{\text{B}} 
   \overset{\Delta G}{\rightleftharpoons}  
   \text{X}_{\text{B}}+\text{X}'_{\text{A}}.
   \label{eq:dg}
\end{equation}
The isotope fractionation free energy $\Delta G$ would be zero if the nuclei behaved 
classically\footnote{Note that in cases where the A and B systems contain a different number of equivalent positions
that can be occupied by the minority isotope (e.g. consider HOD versus CH$_3$D), the equilibrium \eqref{eq:dg}
also involves a component of configurational entropy, that means that \emph{also classically} $\alpha_\text{A-B}\ne 0$.
Here, unless otherwise specified, we will consider the equilibrium between two specified
positions, i.e. we will write $\alpha_\text{A-B}$ to indicate the sole quantum mechanical component 
to fractionation $\alpha_\text{A-B}^\text{tot}/\alpha_\text{A-B}^\text{class}$.}.
Owing to the quantum nature of the X nuclei, instead, it takes on a finite and measurable value which
often depends strongly on the temperature $T$ or the pressure. Isotope fractionation
is usually characterized in terms of the fractionation ratio 
$\alpha_\text{A-B}=\exp(-\beta \Delta G_\text{A-B})$ at the inverse
temperature $\beta=1/k_B T$. 
Since $\alpha_\text{A-B}$ is usually quite close to unity, it is customary to express the 
fractionation ratio in the form $1000\ln(\alpha_\text{A-B})$, and we adopt this convention here.

The isotope fractionation ratio can be determined experimentally with exquisite precision, 
and  is used routinely to gain insight into reactions and phase transformations in 
geochemistry, biology and atmospheric sciences~\cite{werner2001referencing}. 
For instance, isotope fractionation can be used to trace the origin of water in 
clouds~\cite{word+07nat}, or to reconstruct paleoclymatic records~\cite{bern+00sci} 
and even to characterize the origin of wine~\cite{mart+88jafc}.  
When comparing experimentally-determined fractionation ratios with simulations, 
one has to distinguish between equilibrium isotope fractionation -- the ratio that 
is reached after the different phases involved have reached thermodynamic equilibrium, 
and which is due exclusively to the quantum nature of the isotopes involved -- 
and so-called kinetic isotope effects -- that are due to the fact that differences 
in isotope mass affect reaction rates both classically and quantum mechanically, 
and that can lead to accumulation of one isotope in a non-equilibrium process~\cite{vani-mill07jcp}.
Here we will focus on the computational determination of \emph{equilibrium} 
isotope fractionation, so our results can be directly compared with experiments 
only in systems where thermodynamic equilibrium has been reached. 
However, accurate theoretical determination of the equilibrium isotope ratio 
is also useful to set an absolute reference for the equilibrium ratio, 
which can help determining the kinetic factors that contribute to the observed fractionation. 

Evaluating fractionation ratios from simulations is far from trivial, because of the need
of modelling accurately nuclear quantum effects (NQEs). For small gas-phase molecules, one often computes
an approximate quantum mechanical partition function based on harmonic vibrations and 
rigid rotations~\cite{urey47jcs}. However, the harmonic approximations do not allow one to reach an accuracy
on par with the experimental techniques~\cite{ceri-mark13jcp,webb-mill14jpca}, and cannot be applied 
to the condensed phase. Molecular dynamics combined with a 
path integral formalism~\cite{feyn-hibb65book,chan-woly81jcp,parr-rahm84jcp} 
provides an accurate but computationally demanding approach to fully account for the quantum nature
of the nuclei.

The conventional approach to compute the exact fractionation ratio in path integral 
molecular dynamics simulations involves performing a thermodynamic integration of the 
mean quantum kinetic energy of the tagged atom $\left<T\right>$ as a function 
of the isotope mass~\cite{vani-mill07jcp}
\begin{equation}
-\beta \ln(\alpha_\text{A-B})= 
\Delta G_\text{A-B}=\int_m^{m'} \frac{\avg{T}_{\text{B},\mu}}{\mu} 
-\frac{\avg{T}_{\text{A},\mu}}{\mu} \mathrm{d}\mu.
\label{eq:iso-tdfep}
\end{equation}
 Many improvements have been proposed based on this 
scheme, including employing a change of variables to reduce the number of 
thermodynamic integration points, using re-weighting to compute the kinetic energy in 
a substituted system out of a simulation with only the most abundant 
isotope~\cite{ceri-mark13jcp}, and using correlated-noise Langevin dynamics~\cite{ceri-mano12prl}
or high-order path integral factorizations to accelerate
the convergence with the number of replicas~\cite{buch-vani13cpl,mars+14jctc}.
 
Here we demonstrate that it is possible to evaluate the isotope fractionation ratios directly, 
without the need for a thermodynamic integration. The corresponding estimators are derived from
the ratio of the partition functions of the isotope-substituted systems, which is obtained
by a ``virtual substitution'' of the isotopes of the tagged atom. This procedure
is closely related to the the free-energy perturbation estimators for the mass-scaled kinetic
energy introduced in Ref.~\citenum{ceri-mark13jcp}. Since it avoids the mass integration 
altogether it is more convenient, computationally advantageous, 
and immune to errors in the integral for the isotope mass. 

The structure of this paper is as follows. We first introduce the theory underlying
these direct fractionation estimators in Section~\ref{sec:theory}, then
discuss their application to the Zundel cation in Section~\ref{sec:apps-1},
 and to hydrocarbons in Section~\ref{sec:apps-2}. Then, we finally draw our conclusions.

\section{Theory and methods\label{sec:theory}}

\subsection{The path integral partition function}

Consider a system of $N$ distinguishable particles described by the Cartesian 
Hamiltonian
\begin{equation}
H = \sum_{i=1}^{N} {p_i^2\over 2m_i}+V(q_1,\ldots,q_{N}),
\end{equation}
and the corresponding quantum mechanical partition function
\begin{equation}
Z = {\rm Tr}\left[e^{-\beta \hat{H}}\right] \label{eq:z-qm}
\end{equation}
with $\beta=1/k_{\rm B}T$. The imaginary-time path integral formalism~\cite{feyn-hibb65book,chan-woly81jcp,parr-rahm84jcp}
makes it possible to map the quantum mechanical partition function~\eqref{eq:z-qm} onto the classical 
partition function
\begin{equation}
Z \simeq Z_P={1\over (2\pi\hbar)^{3NP}}\int d{\bf p}\,\int d{\bf q}\,e^{-\beta_PH_P({\bf p},{\bf q})},
\label{eq:z-pimd}
\end{equation}
where $\beta_P=\beta/P$. 

The classical ring polymer Hamiltonian $H_P({\bf p},{\bf q})$ describes $P$ copies of the physical system
with corresponding particles in adjacent replicas connected by harmonic springs:
\begin{equation}
H_P({\bf p},{\bf q}) = K_P({\bf p})+V_P({\bf q})+S_P({\bf q}),
\label{h-2nd}
\end{equation}
where we introduced the classical kinetic energy
\begin{equation}
K_P({\bf p}) = \sum_{i=1}^{N}\sum_{j=1}^P {[{p}_{i}^{(j)}]^2\over 2m_i}, 
\end{equation}
the sum of the potential energies of every replica
\begin{equation}
V_P({\bf q}) = \sum_{j=1}^P V({q}_{1}^{(j)},\ldots,{q}_{N}^{(j)}),
\end{equation}
and the spring energy
\begin{equation}
S_P({\bf q}) \equiv \sum_{i=1}^{N} S({q}_{i},{m}_{i}) = \sum_{i=1}^{N}\sum_{j=1}^P {1\over 2}m_i\omega_P^2[q_{i}^{(j)}-q_{i}^{(j-1)}]^2.
\end{equation}
Here the harmonic spring frequency is $\omega_P=1/\beta_P\hbar$, ${q}_{i}^{(j)}$ indicates the three dimensional 
Cartesian coordinates for the $j$th replica of 
the $i$th particle, and the cyclic boundary conditions $q_{i}^{(0)}\equiv q_{i}^{(P)}$ are implied. 
The ring-polymer partition function $Z_P$ converges to the correct quantum mechanical result as $P\rightarrow \infty$. 
Note also that the momenta in Eq.~\eqref{eq:z-pimd} are only sampling devices, and can be integrated out trivially. 
In the following we will only consider the configurational ring polymer partition function
\begin{equation}
\bar{Q} = \int d{\bf q}\,e^{-\beta_P \left(S_P({\bf q})+V_P({\bf q}) \right)  },
\end{equation}
where we omitted the $P$ subscript and the constant prefactors for convenience of notation.

\subsection{Direct estimators for isotope fractionation ratios\label{sec:zeta}}

Assume, without loss of generality, that the Cartesian coordinates of an $\text{X}$ atom are denoted by $q_1$. 
The partition function of system $\text{A}$ containing the $\text{X}$ isotope can be expressed as
\begin{equation}
\bar{Q}_\text{A}=\int_\text{A} d{\bf q}\, e^{-\beta_P V_{P,\text{A}}({\bf q})}
e^{-\beta_P \sum_{i=2}^{N_\text{A}} S({q}_{i},{m}_{i})}
e^{-\beta_P S({q}_{1},m)},
\label{eq:z-a}
\end{equation}
where the integral is meant to extend over the configuration space of system $\text{A}$, and the contribution 
from the spring energy of the $\text{X}$ atoms is singled out, taking into account that $m_1=m$. 
$V_{P,\text{A}}$ is the sum of the potential energy function $V_\text{A}$ over the $P$ path integral beads. 
The partition function $\bar{Q}'_\text{A}$ for the system $\text{A}$ where isotope $\text{X}$ has been substituted by $\text{X}'$ 
can be easily obtained by making the change $m\leftarrow m'$.  $\bar{Q}_\text{B}$ and $\bar{Q}'_\text{B}$ 
are the analogous expressions for the system B.

Let us consider the two phases as separated and non-interacting.
This is always the case if one considers isotope substitution between bulk phases for which
the interface between the two subsystems can be neglected, and a more general formulation that
can be used when there is not a clear-cut distinction between the two states is discussed 
in Appendix~\ref{app:fluctional}.
Then, the combined partition function for 
$\text{X}$ in $\text{A}$ and $\text{X}'$ in $\text{B}$ is just $\bar{Q}_\text{A} \bar{Q}'_\text{B}$.
The relative probability of the two isotope configurations -- which is precisely the 
fractionation ratio -- is the ratio of the combined partition functions
\begin{equation}
\alpha_{A-B}=\frac{\bar{Q}'_\text{A} \bar{Q}_\text{B}}{\bar{Q}_\text{A} \bar{Q}'_\text{B}}.
\end{equation}

\paragraph*{Direct thermodynamic estimator for the isotope fractionation ratio.}

Now consider $\bar{Q}'_\text{A}/\bar{Q}_\text{A}$, which reads
\begin{equation}
\frac{
\int_A d{\bf q}\, e^{-\beta_P V_{P}({\bf q})}
e^{-\beta_P \sum_{i=2}^{N} S({q}_{i},{m}_{i})}
e^{-\beta_P S({q}_{1},m')}
}{
\int_A d{\bf q}\, e^{-\beta_P V_{P}({\bf q})}
e^{-\beta_P \sum_{i=2}^{N} S({q}_{i},{m}_{i})}
e^{-\beta_P S({q}_{1},m)}
}.
\label{eq:za-ratio}
\end{equation}
By multiplying and dividing the integrand at the numerator by $e^{-\beta_P S({q}_{1},m)}$, 
one sees that~\eqref{eq:za-ratio} is just the ensemble average of the direct 
thermodynamic  estimator
\begin{equation}
\mathcal{Z}_{m,m'}^\text{TD} \equiv \exp \left[ 
- \frac{1}{2}\beta_P\omega_P^2(m'-m)\sum_{j=1}^P[q_1^{(j)}-q^{(j+1)}_1]^2\right] 
\label{eq:zetatd}
\end{equation}
over a simulation of system $\text{A}$ containing the isotope $\text{X}$ of mass $m$, 
i.e.
\begin{equation}
\frac{\bar{Q}'_\text{A}}{\bar{Q}_\text{A}}
=\left<\mathcal{Z}_{m,m'}^\text{TD} \right>_{\text{A},m}.
\end{equation}
Note that when there are multiple equivalent $\text{X}$ atoms in the system it is possible to
improve the statistical convergence of the estimator by averaging the values of $\mathcal{Z}_{m,m'}^\text{TD}$ 
computed separately for each equivalent atom.

In a similar way, one can obtain the ratio  $\bar{Q}'_\text{B}/\bar{Q}_\text{B}$, 
as an average of $\mathcal{Z}_{m,m'}^\text{TD}$ over a simulation of the system $\text{B}$,
so that ultimately the fractionation ratio can be expressed as
\begin{equation}
\alpha_{A-B}^\text{TD}=
\frac {\left<\mathcal{Z}_{m,m'}^\text{TD} \right>_{\text{A},m}}
{\left<\mathcal{Z}_{m,m'}^\text{TD} \right>_{\text{B},m}}.
\label{eq:alpha-td}
\end{equation}
In order to evaluate the isotope fractionation ratio between systems $\text{A}$ and $\text{B}$, 
only \textit{two} simulations need to be performed, containing only the naturally abundant isotope. 
Furthermore, the computation for each value of $\mathcal{Z}_{m,m'}^\text{TD}$ is inexpensive, as it 
does not involve evaluating the physical potential but just the harmonic spring energy. 

\paragraph*{Connection with the free-energy perturbation estimators of the particle kinetic energy.}
This direct thermodynamic estimator is closely related to the thermodynamic 
free energy perturbed (i.e. re-weighted) (TD-FEP) kinetic energy estimator introduced in Ref.~\citenum{ceri-mark13jcp}.
Consider a ``pure'' thermodynamic version (in Ref.~\citenum{ceri-mark13jcp}
the centroid-virial form $T_\text{CV}(q) = \frac{1}{2\beta} +\frac{1}{2P} \sum_{j=1}^{P} 
           \left(q^{(j)}-
             \overline{q}\right)
           \frac{\partial V} {\partial q^{(j)}}$
was used for the kinetic energy term)
\begin{equation}
\avg{T_\text{TD}}_{\mu}^\text{TD} = \frac{
\avg{T_\text{TD}(\mu,q) \exp\left[- h_\text{TD}(\mu,q) \right] }_m}
{\avg{\exp\left[- h_\text{TD}(\mu,q) \right]}_m} ,
\label{eq:td-fep}
\end{equation}
with
\begin{equation}
T_\text{TD}(\mu,q) = \frac{1}{2\beta_P} 
-\frac{1}{2P}\mu\omega_P^2 \sum_{j=1}^{P} \left(q^{(j)}-q^{(j-1)}\right)^2,
\end{equation} 
and
\begin{equation}
h_\text{TD}(\mu,q)=
\frac{(\mu-m)\beta_P\omega_P^2}{2} \sum_{j=1}^{P} \left(q^{(j)}-q^{(j-1)}\right)^2.
\label{h-td}
\end{equation}
Also, recall that in this context the fractionation ratio is obtained by
performing a thermodynamic integration with respect to mass (Eq.~\eqref{eq:iso-tdfep}).
It is then easy to see that Eq.~\eqref{eq:alpha-td} is equivalent to performing 
analytically the integral with respect to $\mu$ in \eqref{eq:iso-tdfep} -- for instance
by noticing that $-\beta\frac{\partial}{\partial m'} \ln {\left<\mathcal{Z}_{m,m'}^\text{TD} \right>_{\text{A},m}} $
and $\avg{T_\text{TD}}_{\text{A},m'}^\text{TD}/m'$ differ only by an additive constant that cancels out
when one combines the term for system $\text{B}$ in Eq.~\eqref{eq:iso-tdfep}. 

In the light of this connection, one sees that the direct thermodynamic 
estimator~\eqref{eq:zetatd} can be regarded as a simpler, more convenient form
of the  TD-FEP  estimator. It can be used in all the cases in which 
a free-energy perturbation approach is applicable, and avoids altogether
the need of integrating over different values of the scaled mass $\mu$.

As discussed in Ref.~\citenum{ceri-mark13jcp}, the large 
fluctuations in the exponent of the TD-FEP estimator for large changes in mass
(which is effectively the same term appearing in Eq.~\eqref{eq:zetatd}) lead to 
a dramatic degradation of the sampling performance, which can be recognized 
by the appearance of abrupt changes of the cumulative average during the course
of a simulation. Averaging the direct estimator is not fully 
equivalent to a re-weighing procedure,
so the results of Ref.~\citenum{ceri+12prsa} do not strictly apply. Still as we 
will see, the large fluctuations in the exponential, and their
growth with the number of beads $P$ make the use of Eq.~\eqref{eq:zetatd}
impractical at low temperature and/or for isotope mass ratios very different from one. 

\paragraph*{Direct scaled-coordinates estimator for the isotope fractionation ratio.}

To obtain a more favorable behavior of the statistical efficiency with large number
of beads $P$ and isotope mass ratio, we derived an alternative scaled-coordinates estimator $\mathcal{Z}_{m,m'}^{\text{SC}}$. 
We consider again the ratio of partition functions $\bar{Q}'_\text{A}/\bar{Q}_\text{A}$ as in 
Eq.~\eqref{eq:za-ratio}, and we apply the change of variables
\begin{equation}
{q}_1^{(j)}\leftarrow\overline{q}_1 + 
\sqrt{\frac{m'}{m}} \left(q_1^{(j)}
-\overline{q}_1 \right),
\label{eq:chg-q}
\end{equation}
to the integral at the numerator.
Here $\overline{q}_1=\sum_j q^{(j)}_1/P$ is the position of the centroid, and the change
of variables corresponds to a scaling of the ring polymer coordinates relative to the
centroid~\cite{yama05jcp}.
With this transformation, the integral in the numerator of Eq.~\eqref{eq:za-ratio} 
can be written in the form
\begin{multline}
\vert J \vert\int_A d{\bf q}\, e^{-\beta_P V_{P}({\bf q})}
e^{-\beta_P \sum_{i=2}^{N} S({q}_{i},{m}_{i})}e^{-\beta_P S({q}_{1},m)}\\
\times e^{-\beta_P \left[ 
\sum_{j=1}^P V({q}_{1}'^{(j)},\ldots,{q}_{N}^{(j)})-V({q}_{1}^{(j)},\ldots,{q}_{N}^{(j)})
\right] }
\end{multline}
where ${q}_{1}'^{(j)}= \overline{q}_1 + \sqrt{\frac{m}{m'}} \left(q_1^{(j)}
-\overline{q}_1 \right)$, and the Jacobian $\vert J \vert$ takes a constant value.
Performing analogous manipulations for $\bar{Q}'_B$ in $\bar{Q}'_B/\bar{Q}_B$, 
one sees that the isotope fractionation ratio can be written as 
\begin{equation}
\alpha_{A-B}^{\text{SC}} = \frac {\left< \mathcal{Z}_{m,m'}^{\text{SC}} \right>_{A,m}}
{\left< \mathcal{Z}_{m,m'}^{\text{SC}} \right>_{B,m}},
\end{equation}
where the Jacobians have cancelled out, and we have introduced a
 direct scaled-coordinates estimator
\begin{equation}
\begin{split}
\mathcal{Z}_{m,m'}^{\text{SC}} \equiv \exp \left[ 
-\beta_P\sum_{j=1}^P V({q}_{1}'^{(j)},\ldots,{q}_{N}^{(j)})-\right. \\
\left. \vphantom{\sum_{j=1}^P} V({q}_{1}^{(j)},\ldots,{q}_{N}^{(j)})\right], 
\end{split}\label{eq:zetasc}
\end{equation}
that can be used in the same way as its $\mathcal{Z}_{m,m'}^{\text{TD}}$ counterpart. 
Each evaluation of Eq.~\eqref{eq:zetasc} requires computing the inter-atomic
potential, so this estimator is not as inexpensive as the thermodynamic version. 
At low temperature or for large isotope mass ratios, however, the gain in statistical
efficiency more than offsets the additional computational cost. 

It is worth mentioning that also this estimator can be shown to correspond to an exact,
analytical mass integration of the scaled-coordinates free energy perturbed kinetic energy estimator (i.e. SC-FEP)  
introduced in Ref.~\citenum{ceri-mark13jcp}. Contrary to the case of $\mathcal{Z}^{\text{TD}}_{m,m'}$,
performing the integration analytically is more than a matter of practical convenience. 
Each evaluation of the SC-FEP estimator
requires a separate calculation of the potential energy function, and so avoiding the thermodynamic
integration not only reduces the computational cost, but also circumvents
discretization error even in cases in which the smoothing of the integrand introduced in 
Ref.~\citenum{ceri-mark13jcp} is not effective. 
Furthermore, the evaluation of $\mathcal{Z}_{m,m'}^{\text{SC}}$ does not require computing
derivatives of $V$, which might be convenient in cases where the forces are not used for 
sampling, as it is the case in Monte Carlo path integral simulations.

In the above discussion, $\mathcal{Z}_{m,m'}^{\text{SC}}$ is taken from the simulations 
of the most abundant isotope type, which seems like the obvious choice. However, in 
some circumstances, it might be advantageous to perform the backward substitution,
by sampling from a simulation containing one atom of the rare isotope $\text{X}'$.
As we will comment on in Section~\ref{sec:apps-1}, this is due to the fact that
$\mathcal{Z}_{m',m}^{\text{SC}}$ has better statistical properties for $m'>m$.

The functional form of the direct estimators described above is based on the 
primitive second-order Trotter expansion of the Boltzmann operator. 
The generalization to higher-order expansions is straightforward. 
The derivation of the direct estimators under the Takahashi--Imada scheme~\cite{takahashi1984monte} 
is discussed in the Appendix~\ref{sec:zeta-ti}. As we will see in practical cases,
applying the fourth-order correction may accelerate the convergence with respect to the number 
of beads $P$, but may also impact adversely the statistical efficiency of the estimators.

Another promising approach to reduce the number of path integral replicas needed 
for convergence is that of using correlated-noise Langevin dynamics~\cite{ceri-mano12prl}.
In its current form, however, the path integral Generalized Langevin Equation (PIGLET) thermostat
does not enforce the bead-bead correlations that are necessary to accelerate the convergence of 
complex estimators such as those introduced in the present work. 
In the specific case of the liquid-vapor fractionation in water, a coincidental cancellation of errors
makes it possible to use free-energy perturbation estimators together with PIGLET~\cite{ceri-mark13jcp,wang+14jcp}.
Unfortunately, in all the examples we tested here results were not reliable, so using this option should
only be considered after careful testing. It is however worth mentioning 
that, by computing inexpensively the kinetic energy of isotopes, PIGLET techniques can be 
very useful when one prefers to use conventional thermodynamic integration, or when one simply wants to estimate
roughly $\alpha_\text{A-B}$ based on the difference in kinetic energy between the two phases.

\section{Position-specific and clumped isotope ratios in $\text{H}_5\text{O}_2^+$~\label{sec:apps-1}}

The Zundel cation $\text{H}_5\text{O}_2^+$, together with the Eigen cation $\text{H}_3\text{O}^+$, is one of the 
limiting states that are used to rationalize the behavior of a solvated proton, and plays a crucial role 
in our understanding of proton transport~\cite{marx+99nat} and the autoionization of water~\cite{hass+11pnas}.
Extensive experimental~\cite{YehLee1989, GuascoJohnson2011, AsmisScience2003, HammerBowmanCarter2005, FridgenMaitre2004} 
and theoretical~\cite{AgostiniCiccotti2011, ParkKim2007, VenerSauer2001, SauerDoebler2005, ChengKrause1997, VendrellMeyer2007, 
BaerMarxMathias2010, KaledinBowmanJordan2009, HuangBraamsBowman2005} 
 studies of an isolated Zundel cation in the gas phase have been conducted, most notably
by means of vibrational spectroscopy~\cite{Stoyanov2010,Olesen2011},
to probe its structure and the extent of the coupling between its constituent atoms.
In both experimental and computational studies, the H/D isotopic substitution were found to significantly 
affect the intensity of infrared peaks of mixed H/D species~\cite{Devlin2005,McCunn2008}. 
One observes a clear preference for H to occupy the central position of the singly-deuterated Zundel ion
-- which can be qualitatively understood as the configuration that minimizes the total zero-point energy 
of the substituted molecule.

\paragraph*{Simulation details}
We have evaluated quantitatively the position-dependent H/D fractionation 
ratio between the central and peripheral positions in the singly-deuterated Zundel cation
$\text{H}_4\text{D}\text{O}_2^+$, as well as the clumped-isotope $^{16}$O/$^{18}$O enrichment,
which measures the propensity of rare isotopes to clump together. We used a high-accuracy 
potential energy surface fitted from CCSD(T) data~\cite{HuangBraamsBowman2005},
so our results should represent quantitative predictions that can be verified
by spectroscopic measurements. 
We performed path integral molecular dynamics (PIMD) simulations at 300K, 
using the implementation of PIMD and of the direct fractionation estimators which
are available in the i-PI code~\cite{ceri+14cpc}. We used an efficient local 
path integral Langevin (PILE) thermostat with 
targeted optimal sampling on the internal ring-polymer normal modes ($\gamma_k=\omega_k$) 
and a white-noise Langevin thermostat applied to the centroid, with a time constant of 
1ps~\cite{ceri+10jcp}.  A time step of 0.25fs was used, and each simulation was run for 
$2\times10^6$ steps including the equilibration of $20,000$ steps. We computed and output every
four steps the direct estimators ($\mathcal{Z}_{m,m'}^{\text{TD}}$ and 
$\mathcal{Z}_{m,m'}^{\text{SC}}$) along with their fourth-order counterparts 
($\mathcal{Z}_{m,m'}^{\text{TD-TI}}$ and $\mathcal{Z}_{m,m'}^{\text{SC-TI}}$, 
see Appendix~\ref{sec:zeta-ti}). In all cases we computed the estimators 
every 4 time steps and for a single 
instance of each atom kind, so as to be able to compare statistical efficiency in a 
straightforward manner. As we will discuss later, the efficiency of the TD estimator
could be improved inexpensively by evaluating it more often and for all the 
statistically-equivalent atoms in the simulation. 

\begin{figure}
\centering
\includegraphics{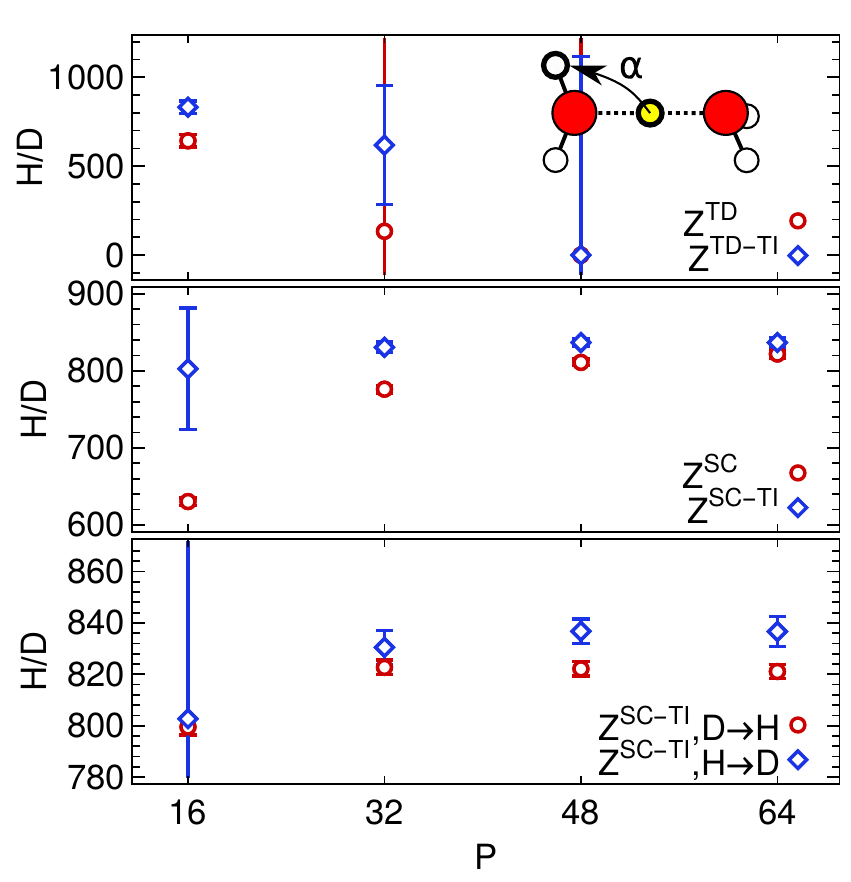}
    \caption{The position-specific isotope ratio $1000\ln\alpha$ of H/D between 
a peripheral position and the central position in 
${\text{H}_2\text{O}}\cdot{\text{H}^+}\cdot{\text{OH}_2}$, as a function of the number of beads.
Red, white and yellow spheres indicate O, H and D, respectively. The top panel reports the 
results for the thermodynamic estimator, the middle panel reports the scaled-coordinates
estimators, and the bottom panel compares the results for forward H$\rightarrow$D 
and backwards D$\rightarrow$H substitution.
}
    \label{fig:f1}
\end{figure}

\paragraph*{Convergence and efficiency of the direct estimators for $\text{H}_5\text{O}_2^+$}

Let us examine the convergence of the isotope fractionation ratios 
with respect to the number of beads for the Zundel ion, and compare critically
the relative statistical efficiency of the various estimators that have been 
introduced in the present paper.
Fig.~\ref{fig:f1} reports the isotope fractionation ratio for deuterium to substitute hydrogen
at one of the peripheral positions comparing to the central ``charged'' position, 
with the equilibrium 
\begin{equation}
{\text{H}_2\text{O}}\cdot{\text{D}^+}\cdot{\text{OH}_2}
\overset{\Delta G}{\rightleftharpoons}
{\text{DH}\text{O}}\cdot{\text{H}^+}\cdot{\text{OH}_2}
\label{eq:h-zundel}
.\end{equation}
The upper panel uses the direct thermodynamic estimator~\eqref{eq:alpha-td}, whose fluctuations increase with the number 
of beads. For any number of beads greater than 16, the statistical error on $\mathcal{Z}_{m,m'}^{\text{TD}}$ 
is huge, making this estimator useless for determining H/D fractionation at room temperature or below.
 At $P=64$, our best estimate using the thermodynamic estimator is completely outside of the range of the figure.

The middle panel reports the results from the direct scaled-coordinates estimator $\mathcal{Z}_{m,m'}^{\text{SC}}$. 
In analogy with the centroid virial kinetic energy estimator $T_\text{CV}$~\cite{yama05jcp} 
and the SC-FEP estimator~\cite{ceri-mark13jcp}, the fluctuations are independent on the 
number of beads. Using the primitive second-order Trotter formulation, 48 beads are 
needed to reduce the systematic error to about 3\% in $1000\ln\alpha$. 
The fourth-order Takahashi--Imada estimator ($\mathcal{Z}_{m,m'}^{\text{SC-TI}}$, Eq.~\eqref{eq:zetasc-ti}), 
accelerates substantially the convergence with respect to the number of beads, at the expense however 
of a larger statistical uncertainty for low numbers of beads due to the effect of the reweighing procedure~\cite{ceri+12prsa}. 

The calculations discussed above were performed by computing direct estimators in a simulations of the
 $\text{H}_5\text{O}_2^+$ molecule, effectively performing the virtual substitution $\text{H}\rightarrow\text{D}$
for the central hydrogen atom, and one of the peripheral atoms. It is however also possible 
to evaluate the same quantities performing simulations of 
$\text{DHO}\cdot{\text{H}^+}\cdot{\text{OH}_2}$ and ${\text{H}_2\text{O}}\cdot{\text{D}^+}\cdot{\text{OH}_2}$,
and computing the direct estimators for the virtual backward substitution $\text{D}\rightarrow\text{H}$.
The lower panel of Fig.~\ref{fig:f1} illustrates such results, which are affected by a considerably lower 
statistical error compared with the forward exchange $\text{H}\rightarrow\text{D}$. 

\begin{figure}
\centering
\includegraphics{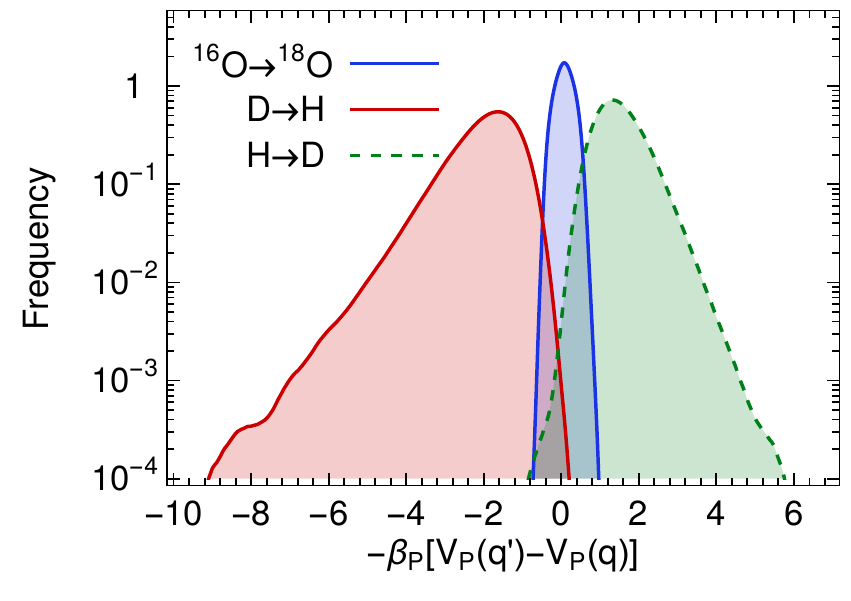}
    \caption{The frequency count for the argument of the exponential term entering 
the $\mathcal{Z}_{m,m'}^{\text{SC}}$ estimator. The forward and backward H/D substitutions
are compared, as well as the exchange of the two oxygen isotopes.}
    \label{fig:f2}
\end{figure}

It is worth investigating in some detail for the reason why the substitution from higher to lower mass
has better statistical properties. Following the discussion about the efficiency of the 
scaled-coordinates free-energy perturbation estimator in Ref.~\citenum{ceri-mark13jcp}, 
the opposite is expected: the fluctuations of the difference Hamiltonian $-\beta_P\left[V_P(q')-V_P(q)\right]$ 
are predicted to be larger for $\text{D}\rightarrow\text{H}$ than for $\text{H}\rightarrow\text{D}$ substitution.
As shown in Figure~\ref{fig:f2}, the distribution of the weights is indeed broader for the 
backward substitution. However, the weight distribution for $\text{D}\rightarrow\text{H}$ has 
a long non-Gaussian tail for negative values of $-\beta_P\left[V_P(q')-V_P(q)\right]$
so that a large fraction of configurations effectively contributes next to nothing to the accumulated average.
Conversely, the distribution for the $\text{H}\rightarrow\text{D}$ case has a strongly non-Gaussian
tail for \emph{positive} values of $-\beta_P\left[V_P(q')-V_P(q)\right]$. Outliers in this direction will 
yield extremely large contributions to the average, despite occurring with very low probability.
Each time a new outlier is encountered, it will dominate and shift dramatically the 
value of the cumulative average, which shows a characteristic zig-zag pattern and a slowly decaying 
statistical error. 
These observations highlight the importance of using extreme care when averaging exponentials, or performing
re-weighing averages. While an analysis based on a Gaussian ansatz for the argument of the exponential
allows one to assess qualitatively the ``comfort zone'' of applicability of a method~\cite{ceri+12prsa}, 
deviations from Gaussian statistics can have a dramatic impact on the quantitative accuracy
of these predictions. 

\begin{figure}
\centering
\includegraphics{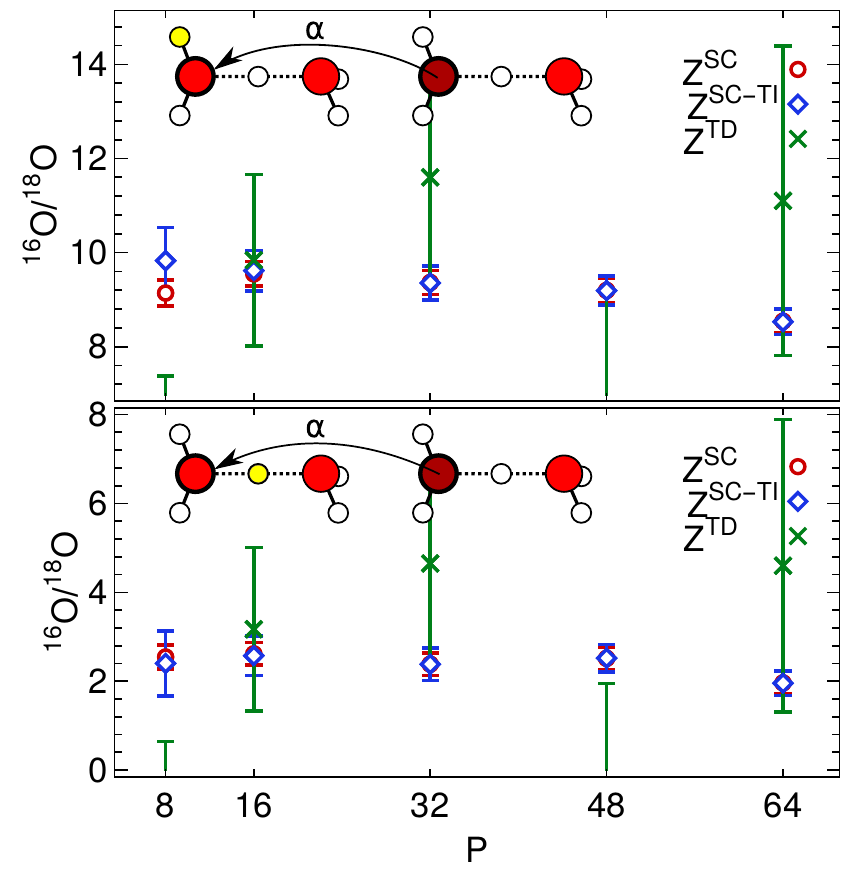}
    \caption{The clumped isotope ratio $1000\ln\alpha$ of $^{16}\text{O}/^{18}\text{O}$ in the deuterated Zundel ion, as a function of the number of beads. The upper panel refers to the clumped ratio with 
${\text{DH}\text{O}}\cdot{\text{H}^+}\cdot{\text{OH}_2}$, while the lower panel considers the case of 
${\text{H}_2\text{O}}\cdot{\text{D}^+}\cdot{\text{OH}_2}$. 
$^{18}\text{O}$ is indicated by darker color, and D by bright yellow.}
    \label{fig:f3}
\end{figure}

Simulations with the isotope-substituted species also made it possible to evaluate
the clumped-isotope fractionation ratio for $^{16}\text{O}/^{18}\text{O}$. This corresponds to the 
enhanced probability of finding a rare isotope $^{18}\text{O}$ in a molecule that also 
contains deuterium rather than in $\text{H}_5\text{O}_2^+$, corresponding e.g. to the equilibrium
\begin{equation}
\begin{split}
{\text{DH}\text{O}}\cdot{\text{H}^+}\cdot{\text{OH}_2} + 
{\text{H}_2(^{18}\text{O})}\cdot{\text{H}^+}\cdot{\text{OH}_2} 
\overset{\Delta G}{\rightleftharpoons} \\
{\text{H}_2\text{O}}\cdot{\text{H}^+}\cdot{\text{OH}_2} + 
{\text{DH}(^{18}\text{O})}\cdot{\text{H}^+}\cdot{\text{OH}_2}.
\end{split}\label{eq:clumped-zundel}
\end{equation}
Fig.~\ref{fig:f3} shows the clumped isotope fractionation ratio corresponding to  
the equilibrium in Eq.~\eqref{eq:clumped-zundel}, and the one
corresponding to the alternative deuterium-substituted molecule
${\text{H}_2\text{O}}\cdot{\text{D}^+}\cdot{\text{OH}_2}$.
The ratio converges with number of beads more rapidly than for H/D, and 
has a considerably smaller statistical uncertainly.  This can be easily 
explained considering the mass ratio associated
$\text{m}(^{18}\text{O})/\text{m}(^{16}\text{O})=1.1253$, which is much 
closer to unity than that of the H/D substitution.  
In the language of the free energy perturbation theory, the substitution between atoms 
with similar masses introduce a smaller perturbation, 
so that the original and the mutated systems have a larger overlap in phase space. 
For this reason, the statistical error on $\mathcal{Z}_{m,m'}^{\text{TD}}$
is not as unmanageable as in the case for H/D.
Still, $\mathcal{Z}_{m,m'}^{\text{SC}}$ delivers a considerably better performance.
As shown in Fig.~\ref{fig:f2}, the bandwidth of the exponential term in $\mathcal{Z}_{m,m'}^{\text{SC}}$  
is small, and so in the case of oxygen it is not necessary to resort to a  
backward substitution from the less abundant to the abundant isotope type.

The above analysis highlights the importance of considering carefully the statistical
efficiency of different estimators when performing isotope-substitution simulations.
For instance, we found that the statistical accuracy of thermodynamic estimators is catastrophic in
the case of lighter atoms. The (relatively small) overhead of computing the scaled-coordinates 
estimator is more than compensated by the increase in statistical efficiency -- particularly for
small gas-phase molecules where one cannot exploit horizontal statistics. 
For the same reason, in the conditions that we considered in this example, it is more 
efficient to perform the backward $\text{D}\rightarrow\text{H}$ virtual substitution, 
while the in the case of heavier nuclei one should be able to use the more obvious 
substitution from the abundant to the rare isotope. 
For heavy isotopes, particularly when multiple equivalent atoms are present, or at high temperature,
the thermodynamic estimator could become a more practical option. The discussion 
in Ref.~\citenum{ceri-mark13jcp} can give some guidelines as to whether one can use the 
thermodynamic estimator, or the scaled-coordinate version is to be preferred. Note 
however that, as we have shown, that analysis is only qualitative, because it disregards
the non-Gaussian distribution of the scaled-coordinates difference Hamiltonian.

The Takahashi-Imada estimator $\mathcal{Z}_{m,m'}^{\text{SC-TI}}$ accelerates considerably the 
convergence with number of beads, without significantly affecting the sampling efficiency except for 
$P\le 16$. With this choice, together with the $\text{D}\rightarrow\text{H}$ substitution, and considering the 
most converged simulation with $P=64$, we computed the position-dependent H/D fractionation ratio
$1000\ln\alpha$ to be $821\pm3$ at $300K$.
This means that the probability of finding a D atom at one of the peripheral sites
is enhanced quantum-mechanically by a factor of $2.273\pm 0.006$
compared to the central position. 
Considering also that there are four equivalent peripheral sites and just one central site,
there is a more than 90\% chance that a $\text{DH}_4\text{O}_2^+$ molecule will be observed in the
${\text{DH}\text{O}}\cdot{\text{H}^+}\cdot{\text{OH}_2}$ configuration rather than as
${\text{H}_2\text{O}}\cdot{\text{D}^+}\cdot{\text{OH}_2}$. 
Since we based our calculation  on a very high-accuracy potential energy 
surface~\cite{HuangBraamsBowman2005}, we believe it should be possible to verify 
quantitatively this prediction by measuring experimentally the relative populations.

Using $\mathcal{Z}_{m,m'}^{\text{SC-TI}}$ and $P=64$, 
we also predicted the clumped isotope fractionation ratios $1000\ln\alpha$ for 
$^{16}\text{O}/^{18}\text{O}$ relative to the pure 
$\text{H}_5\text{O}_2^+$ to be $8.5\pm0.3$ 
in ${\text{DHO}}\cdot{\text{H}^+}\cdot{\text{OH}_2}$ and $2.0\pm0.3$ in 
 ${\text{H}_2\text{O}}\cdot{\text{D}^+}\cdot{\text{OH}_2}$. 
Since the clumped isotope enrichment can also be determined~\cite{Eiler2007} by 
mass spectrometry of isotopologues, these theoretical predictions are also
amenable to experimental verification.

\section{Isotope fractionation ratios in hydrocarbons~\label{sec:apps-2}}

The isotopic compositions of gaseous hydrocarbons are used routinely as indicators 
of their origin, maturity and generation mechanism~\cite{Craig1953,Clayton1991}. 
Besides, isotope ratios for hydrocarbons are used as a tool to monitor 
biodegradation~\cite{Morasch2001}, biosynthesis~\cite{Sessions1999} and photosythesis~\cite{Farguhar}. 
Furthermore, they provide evidence for the reconstruction of the ancient biogeochemical 
history~\cite{Hayes1990,PMID:11536462}. In most cases, the accumulation of different
carbon isotopes in plants is due to kinetic isotope effects, and so
the ratios computed here -- that only account for the equilibrium contribution --
cannot be directly compared with observed ratios, but only provide a reference
``baseline''.

To demonstrate the application of direct estimators in this context, 
we evaluated the hydrogen and carbon fractionation ratios between a few small gas phase hydrocarbon molecules.
We used the reactive empirical potential AIREBO to describe the interaction between the atoms,
since it provides a generally applicable framework that could be easily extended to 
other hydrocarbons~\cite{lind-broi10prb,stua+00jcp}. The simulation details for our PIMD calculations
were identical to those indicated in Section~\ref{sec:apps-1}.

\paragraph*{The convergence of direct estimators for gas-phase hydrocarbons.}

To assess the convergence with respect to the number of beads, and to determine the optimal strategy to apply the 
direct estimators to the gas phase hydrocarbons, we started by evaluating the differential 
fractionation of isotopes between methane and 
ethene at 300K.

The corresponding chemical equilibrium can be written as, e.g.
\begin{equation}
\text{C}\text{H}_3\text{D}+\text{C}_2\text{H}_4
\overset{\Delta G}{\rightleftharpoons}
\text{C}\text{H}_4+\text{C}_2\text{H}_3\text{D}.
\label{eq:ch4-c2h4-h}
\end{equation}

\begin{figure}
\centering
\includegraphics{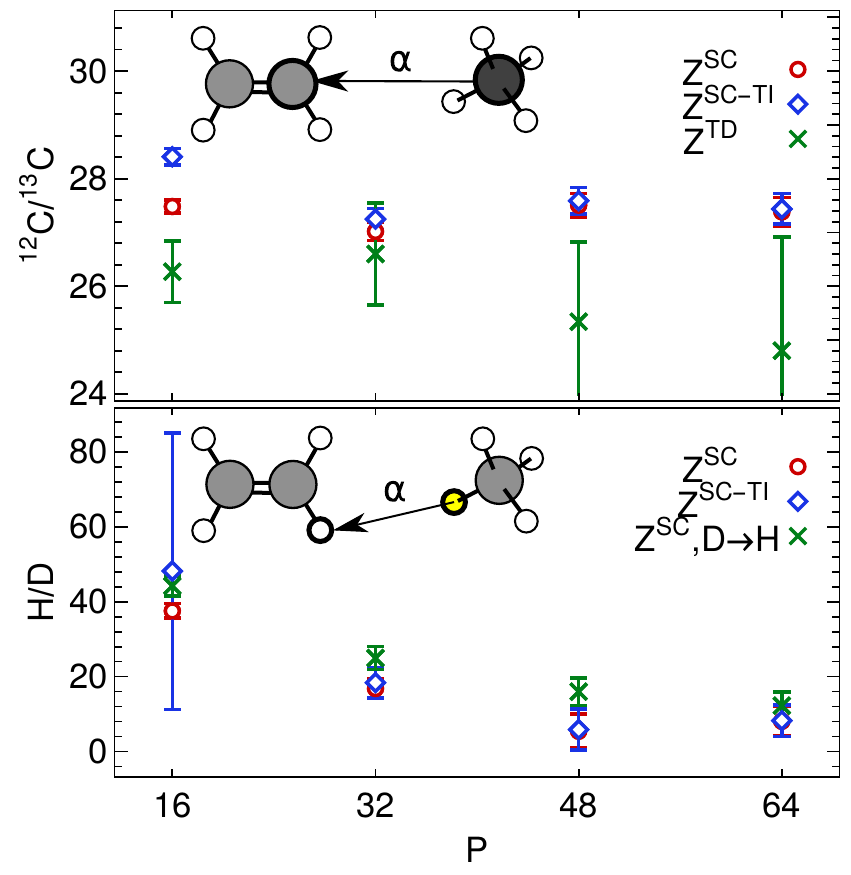}
    \caption{The isotope fractionation ratio $1000\ln\alpha$ for $^{12}\text{C}/^{13}\text{C}$ and $\text{D}/\text{H}$ between $\text{CH}_4$ and $\text{C}_2\text{H}_4$ as a function of the number of beads.
    Grey, dark grey and white spheres indicate $^{12}\text{C}$, $^{13}\text{C}$
    and H, respectively.}
    \label{fig:f5a} 
\end{figure}

Fig.~\ref{fig:f5a} demonstrates the convergence of fractionation ratio of carbon and hydrogen
as a function of the number of beads.
Result for carbon are already almost converged at $P=16$, where the error of $\mathcal{Z}_{m,m'}^{\text{TD}}$ 
is still manageable, as the mass ratio $\text{m}(^{13}\text{C})/\text{m}(^{12}\text{C})=1.083$ 
is very close to one. However, the much smaller error of $\mathcal{Z}_{m,m'}^{\text{SC}}$
means that in the absence of horizontal statistics the scaled-coordinate estimator is still the 
recommended choice. Convergence of H/D fractionation requires at least 48 beads when using the primitive second-order 
Trotter scheme.  As observed in the case of the Zundel cation, 
performing the backward substitution $\text{D}\rightarrow\text{H}$ reduces noticeably the statistical error
compared to the direct $\text{H}\rightarrow\text{D}$ substitution. 
In this case the the fourth-order 
Takahashi--Imada estimator ($\mathcal{Z}_{m,m'}^{\text{SC-TI}}$) does not lead to 
a significant improvement of convergence. As explained in Appendix~\ref{app:fourth},
this is due to a fortuitous cancellation of errors for the second-order estimators.
Note also the very different scales of Fig.~\ref{fig:f5a}
and of Fig.~\ref{fig:f1}: the fractionation ratio in the case of the Zundel cation 
is so large that a difference of 20 units of $1000\ln\alpha$ is a small relative error,
whereas in this case it amounts to 50\%{} of $1000\ln\alpha$. As we will see 
later, H/D fractionation between hydrocarbons involves a considerable degree of 
cancellation between different molecular directions, leading to an overall small
value and apparently to a pronounced cancellation of errors when using Trotter estimators.

\paragraph{Simulation results and discussion}

Based on the convergence tests discussed above, we chose to perform calculations 
with $P=48$, using the scaled-coordinates estimator, and backward substitution to
compute H/D fractionation. We performed simulations at different temperatures
$T_i=\left\{300\text{K},  350\text{K}, 420\text{K}, 510\text{K}, 650\text{K}, 800\text{K}\right\}$, using a parallel-tempering
strategy to improve the statistical convergence~\cite{earl-deem05pccp}. 

We performed calculations for methane ($\text{CH}_4$), 
ethene ($\text{C}_2\text{H}_4$), benzene ($\text{C}_6\text{H}_6$), 
and will report all the fractionation ratios using methane as a reference. 
We also computed the clumped-isotope fractionation ratio for H/D and $^{12}\text{C} /^{13}\text{C}$
in methane, and the position-dependent isotope ratio in propane ($\text{C}_3\text{H}_8$).

\begin{figure}
\centering
\includegraphics{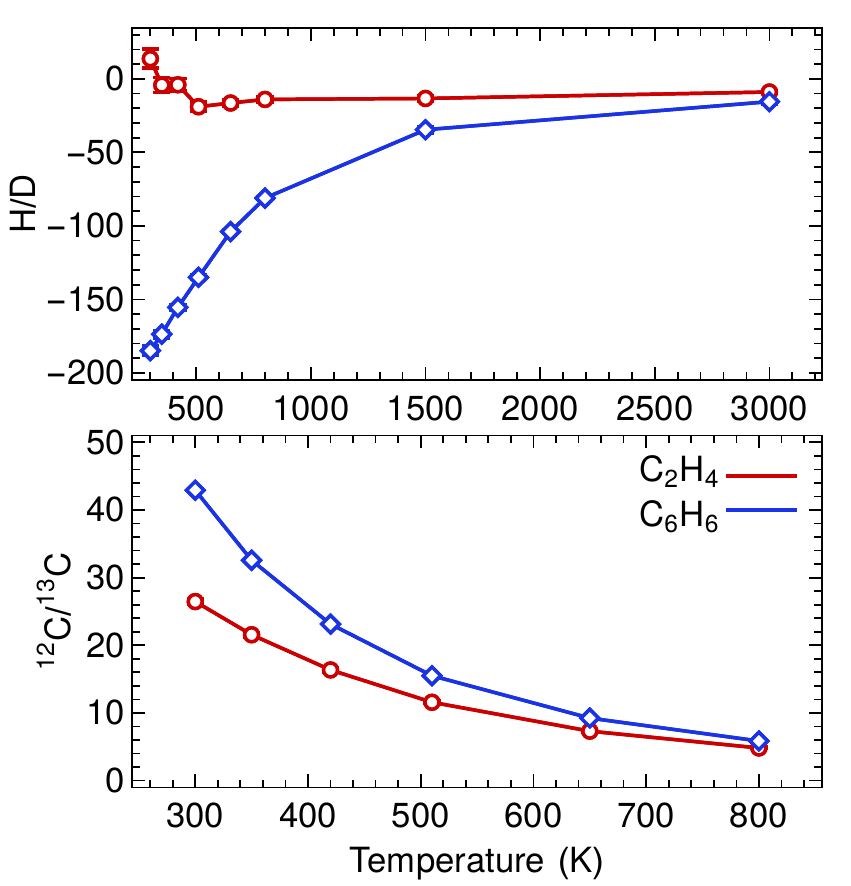}
    \caption{Quantum mechanical component of the H and C fractionation ratios 
between $\text{C}_2\text{H}_4$ and $\text{CH}_4$, and the ratios between $\text{C}_6\text{H}_6$ and $\text{CH}_4$. 
    Notice the difference in the temperature scale.}
    \label{fig:f6a}
\end{figure}

Fig.~\ref{fig:f6a} shows the fractionation ratio for $\text{C}_2\text{H}_4$ and 
$\text{C}_6\text{H}_6$ with reference to $\text{CH}_4$, corresponding to the
chemical equilibrium in Eq.~\eqref{eq:ch4-c2h4-h}.
$^{13}\text{C}$ has a preference for the molecule with the higher number of C--C bonds:
even if carbon-carbon bonds have a lower content of zero-point energy than C--H bonds, the 
carbon atom shares a larger fraction of the zero-point energy, which qualitatively explains
this trend. As the temperature increases, the fractionation ratio decreases
monotonically to zero. 
The trend for H/D fractionation is not as straightforward. On one hand, deuterium has a strong preference 
for $\text{C}\text{H}_4$ relative to $\text{C}_6\text{H}_6$, that only approaches zero at very large 
temperatures. The ethene/methene fractionation ratio is much smaller, and shows an inversion between
a preference for D to reside in ethene at low temperatures to a preference for methane at $T\gtrsim 400K$. 

\begin{figure}
\centering
\includegraphics{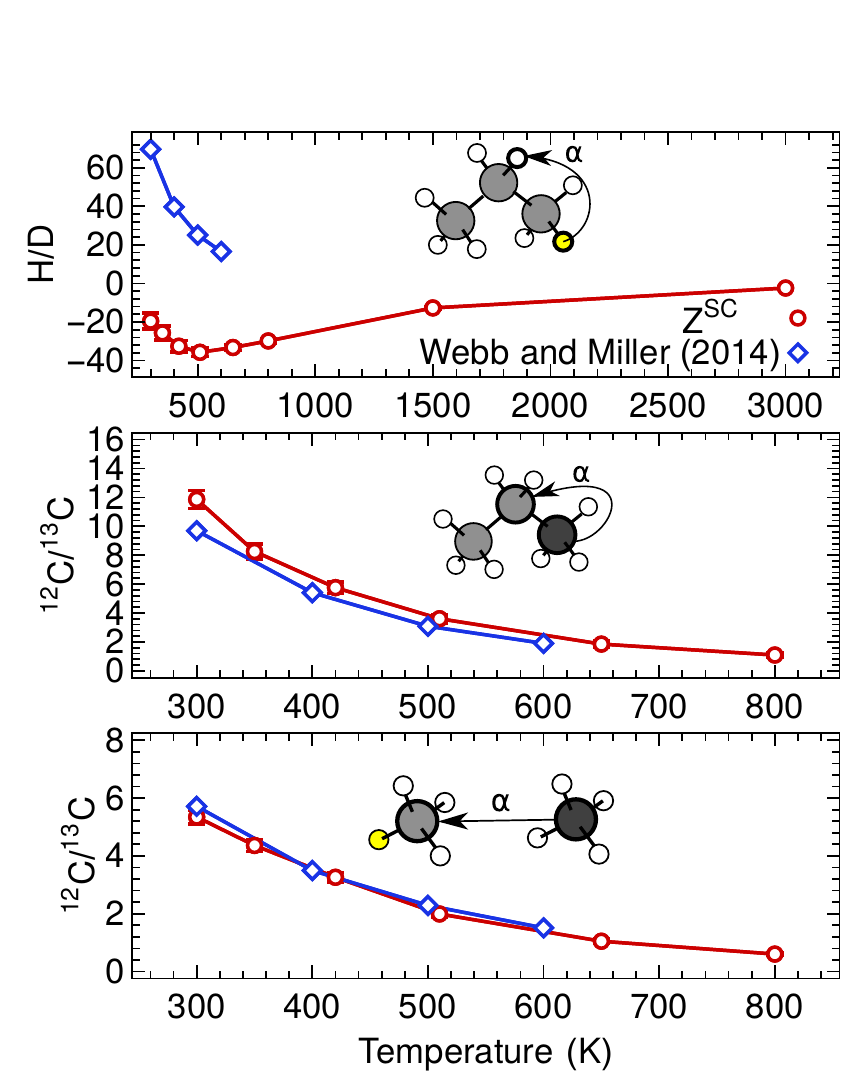}
    \caption{Upper and mid panel: position-specific H and C fractionation ratios in $\text{C}_3\text{H}_8$. 
    Lower panel: clumped-isotopes C fractionation ratio in $\text{CH}_3\text{D}$. 
    Notice the difference in the temperature scale.}
    \label{fig:f6c}
\end{figure}

Before we comment on this inversion, let us consider the case of the position-specific ratios between 
the central CH$_2$ group and the methyl group in propane, for which temperature-dependent results
are shown in the upper and middle panels of Fig.~\ref{fig:f6c}, and that correspond to the equilibria
\begin{equation}
\text{C}\text{H}_3-\text{C}\text{H}_2-\text{C}\text{H}_2\text{D}
\overset{\Delta G}{\rightleftharpoons}
\text{C}\text{H}_3-\text{C}\text{H}\text{D}-\text{C}\text{H}_3
\label{eq:c3h8-h}
\end{equation}
and
\begin{equation}
\text{C}\text{H}_3-{^{12}\text{C}}\text{H}_2-{^{13}}\text{C}\text{H}_3
\overset{\Delta G}{\rightleftharpoons}
\text{C}\text{H}_3-{^{13}\text{C}}\text{H}_2-{^{12}}\text{C}\text{H}_3.
\label{eq:c3h8-c}
\end{equation}
In the lower panel of Fig.~\ref{fig:f6c} we also report
 the clumped-isotope fractionation ratio for $\text{C}\text{H}_3\text{D}$, i.e.
\begin{equation}
{^{13}\text{C}}\text{H}_4+\text{C}\text{H}_3\text{D}
\overset{\Delta G}{\rightleftharpoons}
\text{C}\text{H}_4+ {^{13}}\text{C}\text{H}_3\text{D}.
\label{eq:ch3d-c}
\end{equation}  

The fractionation ratios for these reactions were also discussed in 
Ref.~\citenum{webb-mill14jpca}, where a 
traditional approach based on path-integral Monte Carlo simulations, and 
thermodynamic integration using 5 to 10 integration points were used,
with a total of more than $2\times 10^{10}$ energy evaluations for each 
isotope substitution. The comparison is interesting because of the use 
of a different potential energy surface: the calculations for methane 
in Ref.~\citenum{webb-mill14jpca} were based on a a high-quality potential energy surface
computed at the CCSD(T) level of theory~\cite{lee1995accurate}, whereas 
the simulations for propane were based on the empirical CHARMM PES~\cite{vanommeslaeghe2010charmm}.
Fig.~\ref{fig:f6c} also includes the results from~\citeauthor{webb-mill14jpca} 
to illustrate the comparison.

In the cases involving $^{12}$C$/^{13}$C fractionation, despite the very different potentials we employed, 
our results are both qualitatively and semi-quantitatively consistent with those 
of Ref.~\citenum{webb-mill14jpca}. $^{13}\text{C}$ has a higher affinity for
the central methylene group in propane, and for the deuterated methane -- both 
phenomena that can easily be explained in terms of the partitioning of zero-point 
energy between the atoms in different molecules. 

In the case of H/D fractionation within propane, instead, our calculations
give results that are qualitatively at odds with those of Ref.~\citenum{webb-mill14jpca}.
Results based on the CHARMM force field show preference for deuterium to localize in the methyl groups and
a monothonic dependence on temperature, whereas results based on AIREBO show preference
for methylene, and a non-monothonic dependence that suggests there would probably be an inversion for $T<300$K.
Inversion in fractionation ratios as a function of temperature is observed in the 
liquid/vapor equilibra for water~\cite{mark-bern12pnas,wang+14jcp}, and is consistent with competing quantum
effects between the changes of quantum kinetic energy along different molecular axes. 
For instance, if the change in the frequencies of the stretching and bending modes between 
two molecules have  different sign, the isotope preference might change as the temperature is 
increased, since the effect from the lower-frequency mode will become negligible at lower temperature.
These kinds of cancellations are more likely to occur for light atoms, for which there is larger 
anisotropy between the different components of the kinetic energy. 
It is therefore much harder to capture the qualitative fractionation behavior for light atoms than
it is for the heavier isotopes, which reinforces the idea of using H/D isotope fractionation as 
a sensitive benchmark of different potential energy surfaces~\cite{wang+14jcp}.

The above results show the differential fractionation between a number of gas-phase hydrocarbon 
molecules under thermodynamic equilibrium, and its strong temperature dependence. For natural gas 
there is a general trend for $^{13}\text{C}$ to be depleted in the lighter 
hydrocarbons~\cite{Clayton1991}, which is consistent with our calculations. 
However, because of the complicated chemical and biological processes involved during the 
natural gas production, it is not reasonable to make a direct comparison. 
Theoretical calculations based on more accurate potential energy surfaces can set an absolute 
reference point under the equilibrium conditions, so that the kinetic isotope effects can be singled out 
in the complicated natural gas generation processes.

\section{Discussion and conclusions}

In this paper we have introduced two direct estimators of the isotope fractionation ratio,
that were obtained by computing the ratio of the partition functions of the 
original and isotope-substituted systems. These estimators are closely related
to those that have been recently introduced based on a free energy perturbation
approach~\cite{ceri-mark13jcp}, and therefore their statistical efficiency is largely determined by 
the degree of the phase-space overlap between the original and the substituted systems.
Hence, the statistical efficiency of these direct estimators is more favorable 
when the mass ratio between the isotopes approaches one, or when the system temperature 
increases so that the nuclear quantum effects get smaller.

In both benchmark applications considered here -- the Zundel cation at 300K, and 
gas-phase hydrocarbons at room temperature and above -- the scaled-coordinates version
  gave consistently better performance than the direct thermodynamic estimator,
so that the small additional cost associated with the (relatively infrequent) 
evaluation of $\mathcal{Z}_{m,m'}^{\text{SC}}$
 is more than compensated. On the other hand, $\mathcal{Z}_{m,m'}^{\text{TD}}$ 
can be evaluated almost for free, and so it is advisable to calculate 
it regardless, to provide cross-validation of results.
When there are many equivalent substitution sites in the system,
(e.g. a bulk liquid) it is possible to compute the estimator simultaneously for each site 
and exploit the horizontal statistics. In these cases, particularly for heavy isotopes and 
temperatures above 300K, $\mathcal{Z}_{m,m'}^{\text{TD}}$ 
can be computed for all sites inexpensively, whereas $\mathcal{Z}_{m,m'}^{\text{SC}}$ 
requires a separate potential energy evaluation for each site, which might tip the 
balance in favor of the thermodynamic version of our estimators.

In the case of the H/D substitution, we observed that for $\mathcal{Z}_{m,m'}^{\text{SC}}$
performing the forward substitution from the lighter to the heavier isotope resulted in larger
statistical errors than doing the backwards D$\rightarrow$H substitution.
The analysis in  Ref.~\citenum{ceri-mark13jcp}, which was based on a Gaussian ansatz
for the free-energy perturbation weighing factor, gives the opposite prediction. 
We could explain this discrepancy based on the skewed, asymmetric distribution of 
the exponent in $\mathcal{Z}_{m,m'}^{\text{SC}}$ (Fig.~\ref{fig:f2}).
The implications of this observation are relevant to a much broader class of 
problems than those considered in the present work: similar exponential terms
also enter other methods based on statistical reweighing, such as 
biased molecular dynamics, Jarzynski's inequality, etc.
Considering the shape of the distribution of the exponential term 
can help refine the predictions based on the estimate of its fluctuations, 
and increase the range of applicability of reweighing techniques.

When considering methods to accelerate the convergence of these estimators
we found that unfortunately the combination of path integrals and generalized
Langevin dynamics~\cite{ceri-mano12prl}, that is very effective for the evaluation of NQEs on
structural properties and kinetic energies, does not help in the efficient
evaluations of these estimators, that have a complex non-linear functional 
form and therefore depend on several bead-bead correlations that are not 
currently included in PI+GLE methods. However, we found that 
fourth-order direct estimator ($\mathcal{Z}_{m,m'}^{\text{SC-TI}}$) based on
the Takahashi--Imada factorization can be computed quite easily, and 
they typically lead to improved convergence -- even though in the case
of hydrocarbons a fortuitous cancellation of errors made second-order 
estimators as efficient as those including the TI corrections. 
Since the evaluation of the terms needed in the TI scheme is
completely inexpensive in case where forces are already computed to 
integrate the path integral dynamics, one should always compute them
and assess on a case-by-case basis if the improved convergence
compensates for the statistical errors associated with fourth-order methods. 

The applications we have presented also allowed us to draw some observations
about the isotope-fractionation behavior in the case of the Zundel cation
and of gas-phase hydrocarbons. In H$_4$DO$^+$ one observes a very strong
preference for hydrogen to reside in the central, shared position. These 
findings are consistent with prior experimental and theoretical work, and it 
would be possible to verify them quantitatively. Such a strong fractionation
ratio also suggests that the temperature dependence of $\alpha_\text{A-B}$ 
could be used to assess the temperature of charged water clusters,
 similarly to what was recently proposed for neutral ones~\cite{vide+14jpcl}.
Fractionation in hydrocarbons shows relatively predictable trends in the case
of ${^{12}}$C/${^{13}}$C, with the heavier carbon isotope residing preferentially
in the molecule or position with the largest number of C-C bonds. 
Comparison with previous results showed that this fractionation ratio 
is not very sensitive to the choice of inter-atomic potential. On the contrary,
results for H/D fractionation showed qualitative discrepancies with previous 
work. This discrepancy can be traced to the fact that light elements often
show a strongly anisotropic kinetic energy tensor, with different components that 
can change in opposite directions when changing the atom's environment. 
Since the value of $\alpha_\text{A-B}$ depends on a subtle cancellation 
between different components, it will then be more sensitive to the details
of the potential, and possibly have a non-trivial temperature dependence.

The  direct estimators we have introduced have been implemented in 
the development version of the
path-integral interface i-PI\cite{ceri+14cpc}, making them available for use together 
with a number of empirical forcefields and electronic structure software 
packages, including LAMMPS\cite{plim95jcp}, CP2K\cite{CP2K} and FHI-AIMS\cite{AIMS}. 
Fourth-order corrections based on the Takahashi-Imada scheme can also be 
easily incorporated, accelerating the convergence with respect to the number
of beads. These direct estimators can be regarded as the final step in the process
of making the path integral evaluation of isotope fractionation ratios
easily accessible, by eliminating altogether the need of
performing a thermodynamic integration. They will allow a more
widespread use of computational modelling to provide reference 
equilibrium ratios, and assist the interpretation of fractionation data
in geochemistry, biology and atmospheric sciences.

\appendix
\section{Isotope fractionation in a fluxional setting\label{app:fluctional}}

In the main text we have discussed the case of fractionation of isotopes between distinct phases,
compounds, or molecular sites. Now, let us consider how the direct estimators can be used in a 
setting in which one cannot define a clear-cut distinction between the $\text{A}$ and $\text{B}$
states, but has just one system described by a potential $V$, and the two states are different regions
of configuration space, that are associated to two characteristic functions $\theta_\text{A}(\mathbf{q})$
and $\theta_\text{B}(\mathbf{q})$, which are one if the configuration can be ascribed to one of the states,
and zero otherwise. For instance, the functions could select configurations in which the tagged atom
is at a given distance from an interface~\cite{liu+13jpcc}, or in a specific hydrogen-bonding 
environment. 

The probability of finding the isotope $\text{X}$ in state $\text{A}$ is then given by
\begin{equation}
\begin{split}
\rho_\text{A}=\frac{1}{P\bar{Q}}\int d{\bf q}\, \left[\sum_j\theta_\text{A}(q_1^{(j)},\ldots,q_N^{(j)}) e^{-\beta_P V_{P}({\bf q})}\right. \\
\left. e^{-\beta_P \sum_{i=2}^{N} S({q}_{i},{m}_{i})}
e^{-\beta_P S({q}_{1},m)}\vphantom{\sum_j\theta_\text{A}(q_1^{(j)},\ldots,q_N^{(j)})} \right],
\end{split}
\end{equation}
and the probability of finding the isotope $\text{X}'$ in state $\text{B}$ is
\begin{equation}
\begin{split}
\rho'_\text{B}=\frac{1}{P\bar{Q}'}\int d{\bf q}\, \left[\sum_j\theta_\text{B}(q_1^{(j)},\ldots,q_N^{(j)}) e^{-\beta_P V_{P}({\bf q})}\right. \\
\left. e^{-\beta_P \sum_{i=2}^{N} S({q}_{i},{m}_{i})}
e^{-\beta_P S({q}_{1},m')} \vphantom{\sum_j\theta_\text{B}(q_1^{(j)},\ldots,q_N^{(j)})} \right].
\end{split}
\end{equation}

One quickly sees that the fractionation ratio can be written as $\rho'_\text{A}\rho_\text{B}/\rho_\text{A}\rho'_\text{B}$, so
one is led to evaluate the ratio  $\rho'_\text{A}/\rho_\text{A}$, which can be written as
\begin{equation} 
\rho'_\text{A}/\rho_\text{A}=\left<\theta_\text{A}\mathcal{Z}_{m,m'} \right>_m/\left<\theta_\text{A}\right>_m.
\end{equation}
The ratio can be evaluated by performing a simulation of the abundant isotope $\text{X}$, and 
averaging $\mathcal{Z}_{m,m'}$ selectively over the configurations that are identified as
$\text{A}$ structures. One can compute $\rho'_\text{B}/\rho_\text{B}$ in a similar way,
finally obtaining the desired fractionation ratio.

\section{The fourth order direct estimators using the Takahashi--Imada factorization\label{sec:zeta-ti}}

The convergence of path integral simulations with the number of replicas can be accelerated
by using a high-order factorization of the Boltzmann operator $e^{-\beta \hat{H}}$ rather than 
the simple Trotter factorization that leads to Eq.~\eqref{eq:z-pimd}. Many different
high-order schemes have been proposed, and here we will focus on the Takahashi-Imada
factorization~\cite{takahashi1984monte}, that leads to a partition function with an error
term $\mathcal{O}(\beta_P^4)$, consistent with the Hamiltonian
\begin{equation}
H_{\text{TI}}({\bf p},\textbf{q})=
K_P({\bf p})+V_P({\bf q})+S_P({\bf q})
+V_{\text{TI}}(\textbf{q})
\label{eq:ti-ham}
\end{equation}
which contains a correction term that depends on the squared force
\begin{equation}
V_{\text{TI}}(\textbf{q})=
\frac{1}{24\omega_P^2}
\sum_{i=1}^{N}\sum_{j=1}^P\frac{1}{m_i}
{\left( \dfrac{\partial V({q}_{1}^{(j)},\ldots,{q}_{N}^{(j)})}{\partial q_i^{(j)}}\right)}^2.
\end{equation}
The fourth-order direct estimators can then be obtained as
\begin{equation}
\begin{split}
\mathcal{Z}_{m,m'}^\text{TD-TI} \equiv \exp \left[ 
-\frac{\beta_P}{24\omega_P^2} (\frac{1}{m'}-\frac{1}{m}) 
\sum_{j=1}^P 
{\left( \dfrac{\partial V({q}_{1}^{(j)},\ldots,{q}_{N}^{(j)})}{\partial q_1^{(j)}}\right)}^2 
\right. \\
\left. - \frac{1}{2}\beta_P\omega_P^2(m'-m)\sum_{j=1}^P[q_1^{(j)}-q^{(j+1)}_1]^2\right],
\end{split}
\end{equation}
and
\begin{equation}
\begin{split}
\mathcal{Z}_{m,m'}^{\text{SC-TI}} \equiv \exp \left[ 
\vphantom{\sum_{j=1}^P}
-\beta_P \left( V_{\text{TI}}({q}_{1}',\ldots,{q}_{N})
-V_{\text{TI}}({q}_{1},\ldots,{q}_{N})\right)  
\right. \\
\left. -\beta_P\sum_{j=1}^P V({q}_{1}'^{(j)},\ldots,{q}_{N}^{(j)})-
\vphantom{\sum_{j=1}^P} V({q}_{1}^{(j)},\ldots,{q}_{N}^{(j)})\right].
\end{split}\label{eq:zetasc-ti}
\end{equation}

Performing PIMD based on the Hamiltonian in Eq.~\eqref{eq:ti-ham} would be impractical, 
as the corresponding forces contain terms that depend on the second derivatives of 
the physical potential. In practice, one typically uses a free-energy perturbation
strategy to obtain averages consistent with Eq.~\eqref{eq:ti-ham}, while performing 
sampling based on the second-order partition function in Eq.~\eqref{eq:z-pimd}~\cite{pere-tuck11jcp,ceri+12prsa,mars+14jctc}. 
This entails treating $V_{\text{TI}}$ as a small pertubation to the Trotter Hamiltonian in Eq.~\eqref{h-2nd}, 
and later recovering the fourth-order statistics by re-weighting configurations with the weight 
\begin{equation}
w_{\text{TI}}(\textbf{q})=
{e^{-\beta_PV_{\text{TI}}(\textbf{q})}}
\end{equation}
so that for instance one can compute 
\begin{equation}
\alpha_{A-B}^{\text{SC-TI}} = 
\frac {\left< \mathcal{Z}_{m,m'}^{\text{SC-TI}} w_{\text{TI}}(\textbf{q}) \right>_{(2),A,m} / \left<w_{\text{TI}}(\textbf{q}) \right>_{(2),A,m} }
{\left< \mathcal{Z}_{m,m'}^{\text{SC-TI}}w_{\text{TI}}(\textbf{q}) \right>_{(2),B,m}/\left<w_{\text{TI}}(\textbf{q})\right>_{(2),B,m}  }.
\end{equation}

Note that the performance of reweighed high-order factorizations degrades as the 
size of the system being studied increases~\cite{ceri+12prsa}. However, we noticed
that the Takahashi--Imada factorization yields smaller fluctuations of the 
logarithm of the reweighing factors compared to the Suzuki-Chin factorization that
was used in the analysis in Ref.~\cite{ceri+12prsa}, which can marginally 
extend the range of applicability of this approach. 

\section{Convergence of fourth order estimators for hydrogen isotopes
fractionation between methane and ethene}\label{app:fourth}

\begin{figure}
\centering
\includegraphics{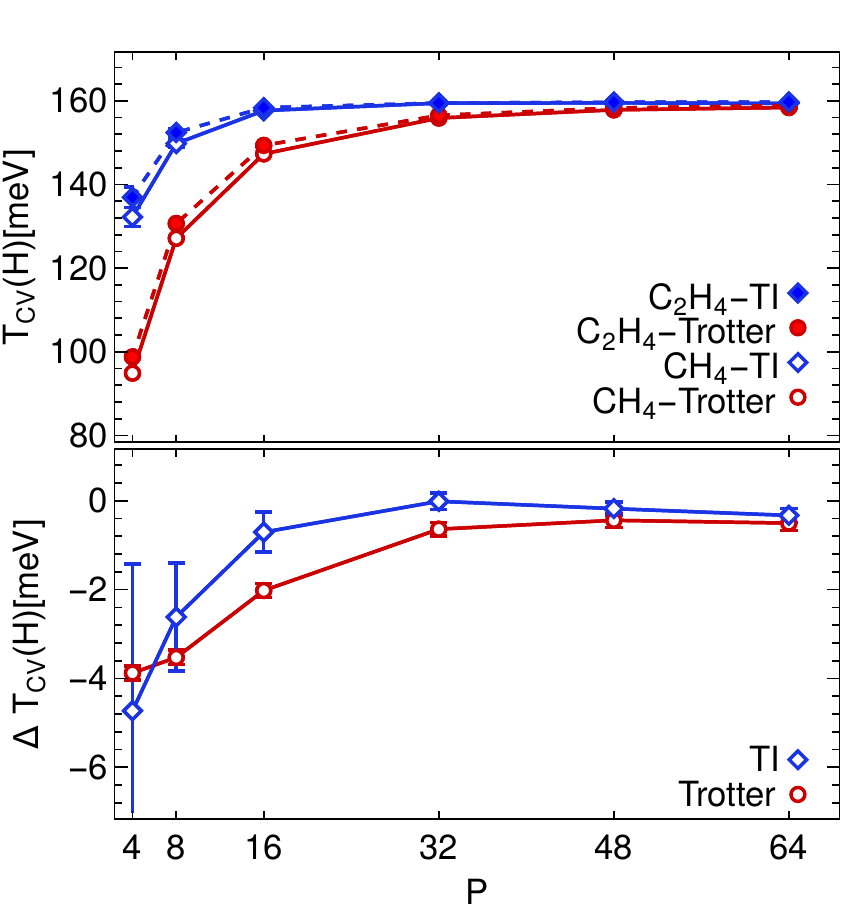}
    \caption{$T_{\text{CV}}$ for $\text{CH}_4$ and $\text{C}_2\text{H}_4$, and $\Delta T_{\text{CV}}=T^{\text{CH}_4}_{\text{CV}}-T^{\text{C}_2\text{H}_4}_{\text{CV}}$, as a function of the number of beads. The lines are provided only as aids for the eye.}
    \label{fig:f5b}
\end{figure}

Contrary to what we observed in the case of H$_5$O$_2^+$,  the fourth-order 
Takahashi--Imada estimator ($\mathcal{Z}_{m,m'}^{\text{SC-TI}}$) does not lead to 
a significant improvement of the convergence with the number of beads. 
To understand why, we computed the effect 
of the Takahashi--Imada correction on the centroid viral kinetic energy\cite{mars+14jctc}
$T_{\text{CV}}$ for $\text{CH}_4$ and $\text{C}_2\text{H}_4$, as well as on their 
difference $\Delta T_{\text{CV}}$ (Fig.~\ref{fig:f5b}) -- which to a first approximation
is proportional to the isotope-substitution free energy~\cite{ceri-mark13jcp}. 
Note that $T_{\text{CV}}$ is far from converged with $P<48$, and the TI 
correction does enhance its convergence considerably. 
However, when it comes to computing $\Delta T_{\text{CV}}$, the errors in the 
Trotter estimate of kinetic energy for the two molecules largely cancel out, 
so that there is little or no improvement when considering the TI correction
to $\Delta T_{\text{CV}}$.

\end{document}